\documentclass{article} 
\usepackage{iclr2026_conference,times}

\usepackage{amsmath,amsfonts,bm}

\def\eqref#1{equation~\ref{#1}}

\def\1{\bm{1}}

\DeclareMathAlphabet{\mathsfit}{\encodingdefault}{\sfdefault}{m}{sl}
\SetMathAlphabet{\mathsfit}{bold}{\encodingdefault}{\sfdefault}{bx}{n}

\usepackage{hyperref}
\usepackage{url}
\usepackage{wrapfig}
\usepackage{graphicx} 
\usepackage{enumitem}
\usepackage{listings}
\usepackage{tcolorbox}
\usepackage{booktabs}
\usepackage{multirow}
\usepackage[table,xcdraw]{xcolor}
\usepackage{longtable}
\usepackage{subfig}
\usepackage{caption}
\usepackage{subcaption}

\usepackage[dvipsnames]{xcolor} 
\usepackage{listings}
\usepackage{float}

\definecolor{DeepLavender}{RGB}{200, 200, 240}
\definecolor{DeepGreen}{RGB}{0, 100, 0}
\definecolor{DeepSkyBlue}{RGB}{0, 120, 200}
\definecolor{VividRed}{RGB}{200, 30, 30}

\lstdefinestyle{agentstyle}{
    backgroundcolor=\color{DeepLavender!40},
    basicstyle=\ttfamily\small,
    breaklines=true,
    columns=fullflexible,
    keepspaces=true,
    frame=single, 
    literate=
      {AGENT'S INTUITION:}{{\bfseries\color{DeepGreen}AGENT'S INTUITION:}}{19}
      {AGENT'S THOUGHT:}{{\bfseries\color{DeepSkyBlue}AGENT'S THOUGHT:}}{16}
      {FINAL ANSWER:}{{\bfseries\color{VividRed}FINAL ANSWER:}}{13}
}

\definecolor{lightgray}{rgb}{0.83,0.83,0.83} 
\lstdefinestyle{promptstyle}{
    basicstyle=\ttfamily\small,
    backgroundcolor=\color{white},
    frame=tb,
    framesep=4pt,
    rulecolor=\color{gray},
    xleftmargin=12pt,
    breaklines=true,
    postbreak=\raisebox{0ex}[0ex][0ex]{\ensuremath{\color{red}\hookrightarrow\space}}
}

\title{AwareCompiler: Agentic Context-Aware Compiler Optimization via a Synergistic Knowledge-Data Driven Framework}

\author{{\bf Hongyu Lin\textsuperscript{\rm 1,2}\thanks{These authors contributed equally to this work.}, Haolin Pan\textsuperscript{\rm 1,2,3}\footnotemark[1], Haoran Luo\textsuperscript{\rm 4,}\thanks{Corresponding author(s).}, Yuchen Li\textsuperscript{\rm 1,2}}, Kaichun Yao\textsuperscript{\rm 2}, Libo Zhang\textsuperscript{\rm 2}, \\ 
{\bf Mingjie Xing\textsuperscript{\rm 2,}\footnotemark[2], Yanjun Wu\textsuperscript{\rm 1,2}} \\
    \textsuperscript{1}University of Chinese Academy of Sciences, China\\
    \textsuperscript{2}Institute of Software Chinese Academy of Sciences, China\\
    \textsuperscript{3}Hangzhou Institute for Advanced Study at UCAS, China \\
    \textsuperscript{4}Nanyang Technological University, Singapore \\
    \texttt{\{hongyu2021,mingjie\}@iscas.ac.cn,} \texttt{haoran.luo@ieee.org} \\
    \texttt{panhaolin21@mails.ucas.ac.cn}   
}

\iclrfinalcopy 
\begin{document}

\maketitle

\begin{abstract}
Compiler optimization is crucial for enhancing program performance by transforming the sequence of optimization passes while maintaining correctness. Despite the promising potential of large language models (LLMs)-based agent for software optimization, automating compiler optimization remains challenging due to: (1) semantic misalignment between abstract program representations and concrete optimization passes, (2) inefficient interaction mechanisms between agents and compiler environments, and (3) reward sparsity from the extensive decision-making process within large optimization spaces. This paper introduces \textbf{AwareCompiler}, an agentic framework for compiler optimization that addresses these challenges through three key innovations: structured knowledge integration and dataset construction, knowledge-driven adaptive pass generation, and data-driven hybrid training pipeline. Experimental results on standard benchmarks demonstrate that AwareCompiler significantly outperforms existing baselines in both performance and efficiency, highlighting the effectiveness of our synergistic knowledge-data-driven approach. Our code is publicly available at~\footnote{Code: \url{https://github.com/LHY-24/AwareCompiler}}.
\end{abstract}

\section{Introduction}
\begin{wrapfigure}{r}{0.6\textwidth} 
  \centering                       
  \vspace{-\intextsep}     
  \includegraphics[width=\linewidth]{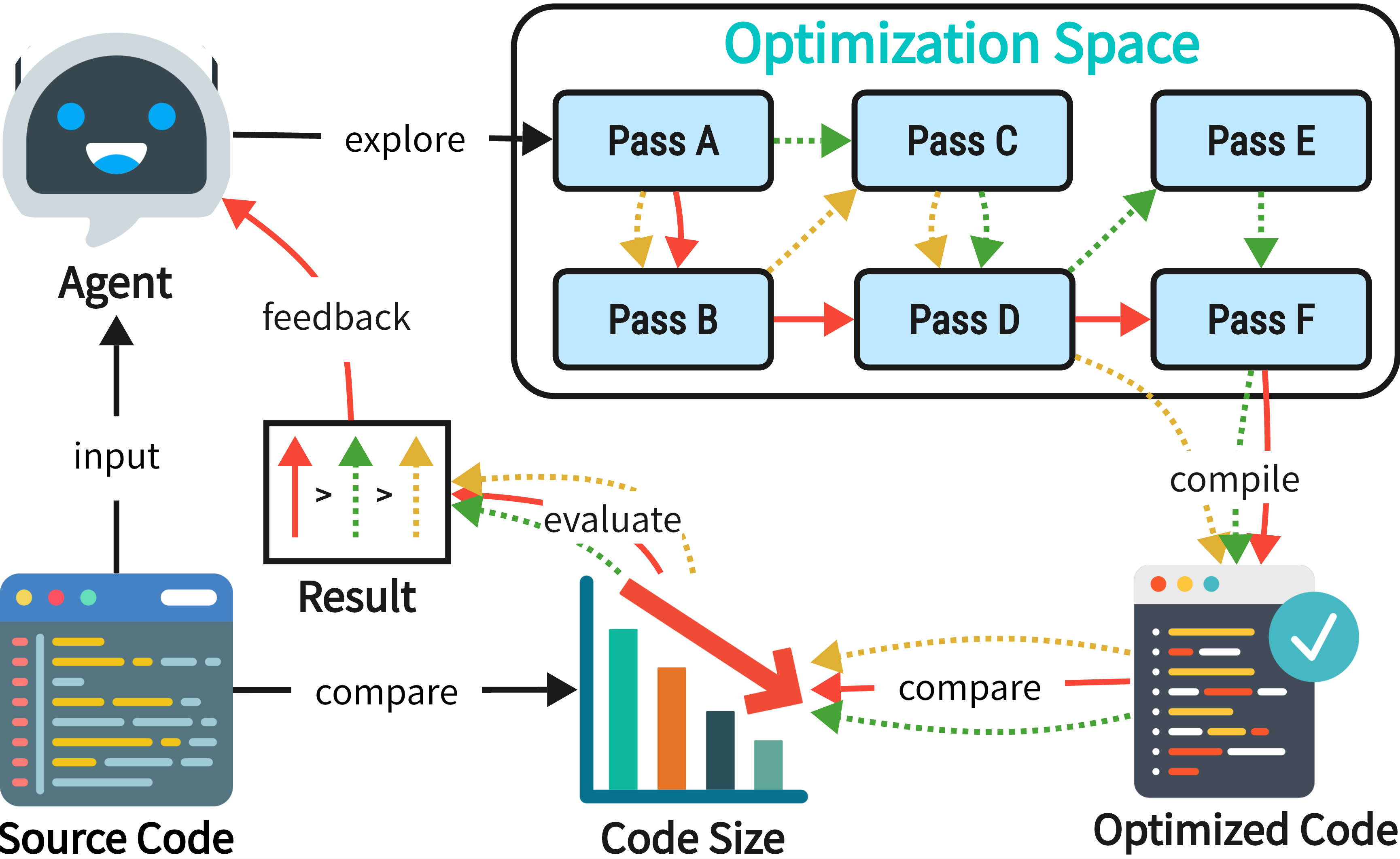} 
  \caption{Overview of compiler optimization task.} 
  \label{fig:task}
  \vspace{-\intextsep} 
\end{wrapfigure}

Compiler optimization is essential for modern computing systems \cite{aho2006compilers}, involving the automated selection and scheduling of optimization passes from a vast space to enhance program performance (see Figure~\ref{fig:task}). Among the various optimization objectives, \textbf{code size} is a key metric \cite{CompilerGym}, as reducing it lowers memory overhead, shortens compilation time, and often improves runtime efficiency. As software systems grow in complexity, the demand for efficient executable code becomes even more critical \cite{gong2025languagemodelscodeoptimization}. 

\begin{figure*}[htbp]
  \centering
  \includegraphics[width=\linewidth]{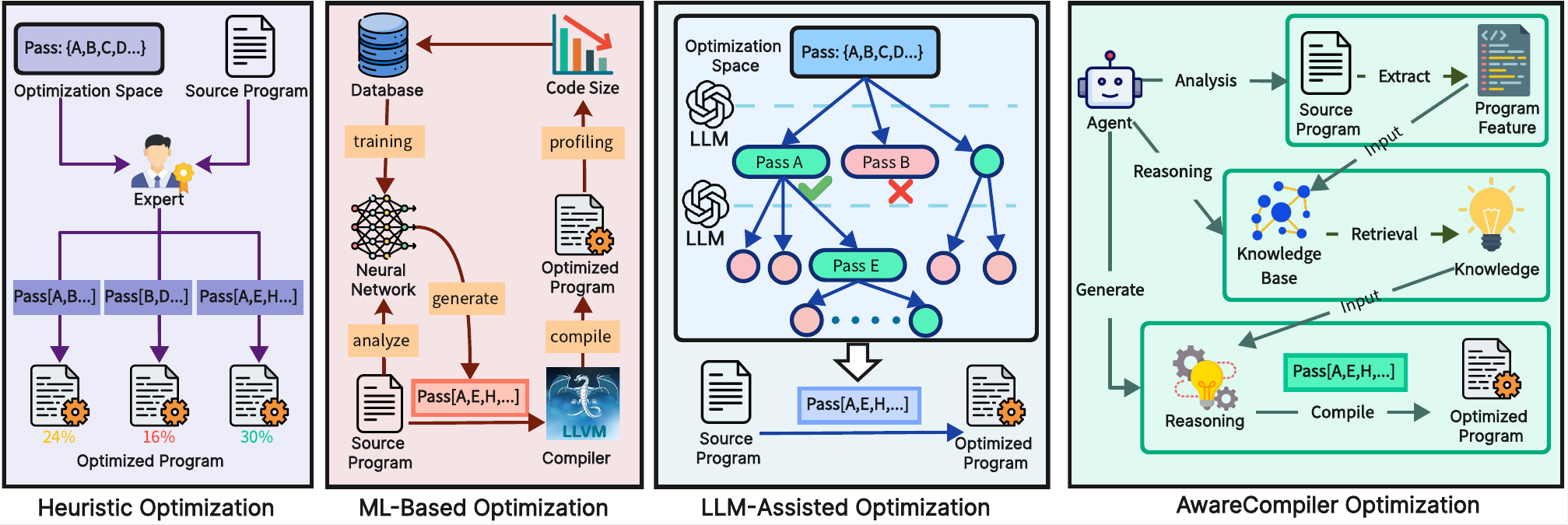}
  \caption{Comparison of different compiler optimization methods, including heuristic-based optimization through the permutations and combinations of "pass", machine learning-based optimization using neural networks, LLM-assisted tuning with hierarchical-based search, and our proposed AwareCompiler method, which incorporates a synergistic knowledge-data driven method.} 
  \label{fig:compare}
\end{figure*}

Historically, compiler optimization has relied on handcrafted heuristics and static cost models for pass selection and scheduling, which were both labor-intensive and time-consuming~\cite{tpe,RIO,opentuner}. With the advent of data-driven methods, machine learning (ML) models have been introduced to automate the selection and ordering of optimization passes~\cite{autophase,BOCA,Comptuner}. While ML-based optimization achieves better adaptivity and performance gains, it suffers from high profiling overhead, extensive compilation requirements, and limited scalability across diverse program domains. Recently, large language model (LLM)-based agents have emerged as a promising new paradigm for compiler optimization~\cite{cummins2024meta,CompilerDream,Compiler-R1}. Leveraging the generalization capabilities of pre-trained models, these agents can understand and generate code across diverse codebases, perform reasoning over program semantics, and even suggest optimization strategies without explicit profiling. This represents a shift from data-specific learning to generalizable knowledge-based reasoning, potentially overcoming the scalability bottlenecks faced by both heuristic and ML-driven approaches.


However, LLM-based agents often produce ineffective or invalid optimization passes due to insufficient contextual reasoning and an inability to predict the real-world effects of optimizations, leading to performance degradation or even program crashes.  Addressing these shortcomings requires tackling three critical challenges: (1) \textbf{Semantic misalignment} between abstract program representations and concrete optimization passes, which leads to plausible but incorrect strategies; (2) \textbf{Ineffective interaction} mechanisms between agents and compiler environments, which typically rely on brute-force exploration; and (3) \textbf{Reward sparsity}, arising from the extensive decision-making horizon in large optimization spaces, preventing effective feedback usage.

This paper introduces \textbf{AwareCompiler}, an LLM-based agentic framework designed to address these challenges. AwareCompiler employs a synergistic knowledge-data-driven approach, empowering agents with agentic context awareness to make context-aware optimization sequences generation. Specifically, AwareCompiler contributes three key innovations: \textbf{(1) Structured Knowledge Integration and Dataset Construction}: We build a symbolic knowledge base that bridges the semantic gap between program representations and pass-level optimizations, along with a tailored high-quality reasoning dataset. \textbf{(2) Knowledge-driven Adaptive Pass Generation}: We propose an adaptive reasoning paradigm that enables agents to analyze code features, retrieve relevant knowledge, and generate optimized sequences, thereby enhancing optimization effectiveness. \textbf{(3) Data-driven Hybrid Training Pipeline}: We model compiler optimization as a multi-turn agent-environment interaction problem and optimize the agent through a hybrid training pipeline incorporating an outcome-based composite reward function, ensuring robust and efficient learning. \par

We evaluate \textbf{AwareCompiler} through extensive experiments on standard benchmarks. The results show that AwareCompiler significantly outperforms existing baselines in code size reduction. Beyond performance improvements, AwareCompiler reduces inconsistent or infeasible optimization passes by incorporating internal reasoning and external knowledge based on program context, ensuring more reliable optimizations. By seamlessly combining knowledge-driven reasoning with data-driven learning, AwareCompiler establishes a solid foundation for next-generation LLM-based compiler optimization agents, paving the way for more efficient and powerful compiler architectures.

\begin{figure*}[htbp]
  \centering
  \includegraphics[width=\linewidth]{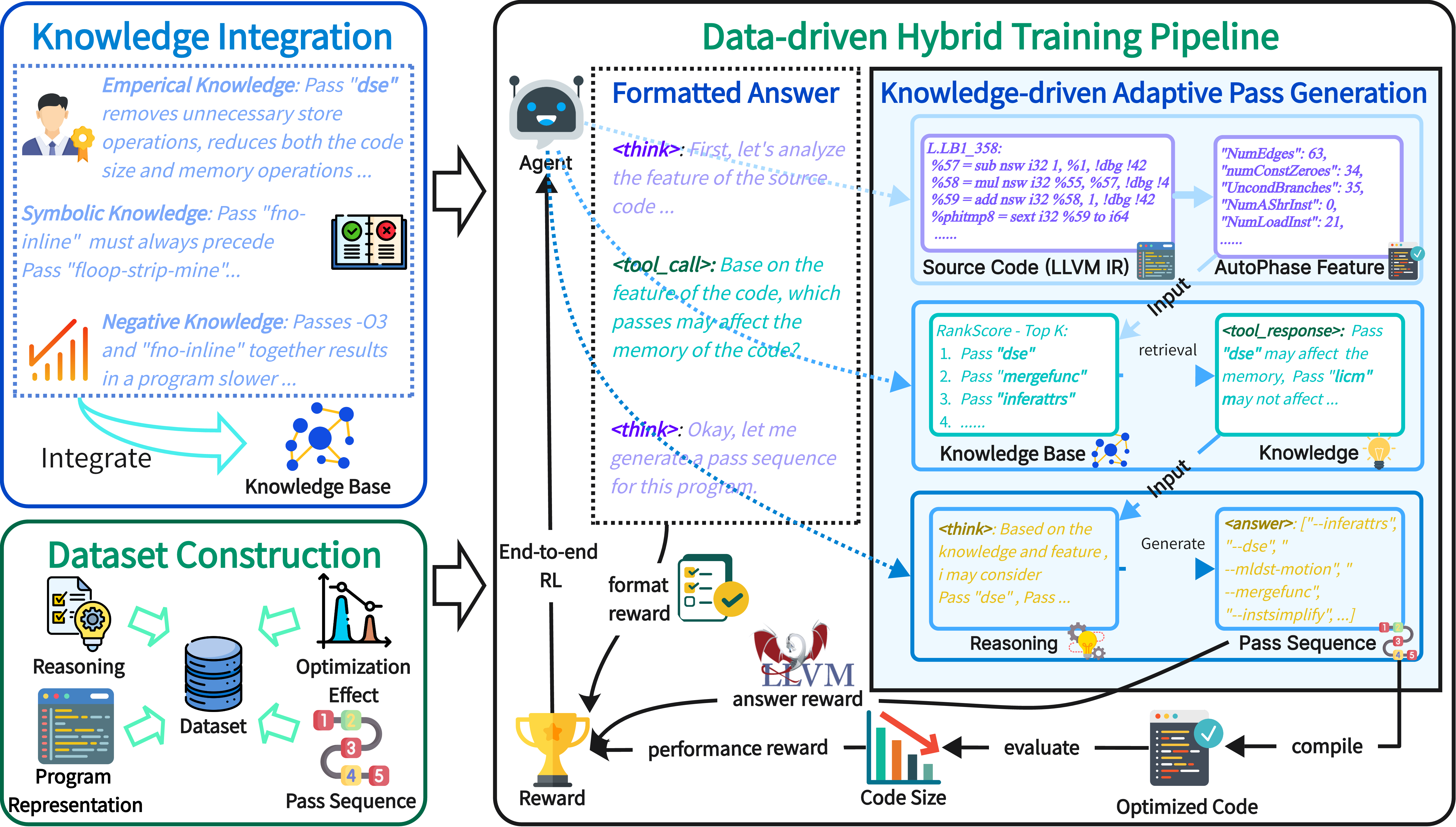}
  \caption{Overview of the framework: AwareCompiler utilizes a synergistic knowledge-data-driven approach to compiler optimization, integrating knowledge-driven adaptive pass generation based on a comprehensive knowledge base (including empirical, symbolic, and negative knowledge) with a data-driven hybrid training pipeline that leverages a curated dataset for model training.}
  \label{fig:framework}
\end{figure*}

\section{Preliminaries}
\subsection{Definition: Optimization Pass and Space}
The optimization space $\mathcal{P} = \{(p_j, \text{sem}(p_j), \text{deps}(p_j), \text{conf}(p_j))\}$ encodes each pass \( p_j \) along with its semantics, dependencies, and conflicts. Here, \( \text{sem}(p_j) \) represents the transformation effect of pass \( p_j \) on the IR, \( \text{deps}(p_j) \) defines the passes that must precede \( p_j \), and \( \text{conf}(p_j) \) denotes the conflicting passes with \( p_j \). An optimization pass sequence is defined as:
\begin{equation}
    \pi = (p_1, p_2, \dots, p_T), \quad p_t \in \mathcal{P}, \quad 1 \leq t \leq T,
\end{equation}
where $T$ is the sequence length. Valid pass sequences must satisfy two constraints: (1) dependency relation: if \( p_i \in \text{deps}(p_j) \), then \( p_i \) precedes \( p_j \); and (2) conflict resolution: if \( p_i \in \text{conf}(p_j) \), then \( p_i \) and \( p_j \) cannot appear together:
\begin{equation}
    \forall p_i, p_j \in \pi, \quad \left( (p_i \in \text{deps}(p_j)) \Rightarrow \text{pos}(p_i) < \text{pos}(p_j) \right) \land \left( (p_i \in \text{conf}(p_j)) \Rightarrow \text{pos}(p_i) \neq \text{pos}(p_j) \right)
\end{equation}
Let $x \in \mathcal{X}$ represent the source code. Applying sequence $\pi$ to $x$ results in the optimized code $x_{\text{opt}} = \mathcal{C}(x, \pi) \in \mathcal{X}$, where $\mathcal{C}$ is the compiler environment and $\mathcal{X}$ is the code space.

\subsection{Problem Formulation: Compiler Optimization}
We formalize compiler optimization for code size reduction as a sequential decision-making problem, where an agent interacts with the compiler environment to iteratively apply passes that transform the program's intermediate representation (IR). The objective of \textbf{code size reduction} can be formalized as $\mathcal{L}_{\text{size}}(x, \pi) = \text{Size}(\mathcal{C}(x, \pi))$, where $\text{Size}(\cdot)$ measures the code size. The agent aims to learn a policy $\pi_\theta: \mathcal{S} \to \mathcal{P}$ parameterized by $\theta$, that minimizes the expected code size while maintaining program correctness:
\begin{equation}
    \pi^*_\theta = \arg\min_{\pi \in \mathcal{P}} \ \mathbb{E}_{\pi_\theta} \big[ \mathcal{L}_{\text{size}}(x, \pi_\theta)],
\end{equation}

\section{Methodology}
In this section, we introduce AwareCompiler, a framework that integrates structured knowledge mapping and dataset construction, data-driven hybrid training pipeline, and knowledge-driven adaptive reasoning paradigm, as illustrated in Figure~\ref{fig:framework}.

\subsection{Structured Knowledge Integration and Dataset Construction}
AwareCompiler leverages a knowledge-data driven method to combine structured compiler knowledge with large-scale training data. Knowledge-driven strategies utilize domain-specific expertise and formal reasoning while data-driven methods excel at identifying patterns in empirical data.

\paragraph{Domain-Specific Knowledge Integration.} The knowledge base $\mathcal{K} = \{\mathcal{K}_{\text{emp}}, \mathcal{K}_{\text{sym}}, \mathcal{K}_{\text{neg}}\}$ encodes compiler-specific domain knowledge, enabling flexible retrieval when the agent’s internal capabilities are insufficient. It is constructed from three primary sources: \\
\textit{(i) Empirical Knowledge}: Captures heuristics and patterns learned from historical optimization data, suggesting optimal pass sequences for code features.
 We represent empirical knowledge as a function $\mathcal{K}_{\text{emp}}$, mapping code features $\mathbf{x}_i$ to optimal pass sequences $\pi_i^*$, expressed as:
\begin{equation}
    \mathcal{K}_{\text{emp}}: \mathcal{X} \to \mathcal{P}, \quad \mathbf{x}_i \mapsto \pi_i^*, \quad \pi^* = \arg\min_{\pi \in \mathcal{P}} \mathbb{E}_{\mathbf{x}_i \sim \mathcal{X}}\big[ \mathcal{L}_{\text{size}}(x, \pi)]
\end{equation}
\textit{(ii) Symbolic Knowledge}: Models the structural relationships between optimization passes, ensuring valid sequences by respecting dependency and conflict constraints, which are captured by the function $\text{deps}(p_j) = \{ p_i \mid p_i \text{ must precede } p_j \}$ and $\text{conf}(p_j) = \{ p_k \mid p_k \text{ conflicts with } p_j \}$:
\begin{equation}
    K_{\text{sym}} = \{ p_j, \text{deps}(p_j), \text{conf}(p_j) \}, \text{deps}(p_j) = \{ p_i \mid p_i \prec p_j \}, \text{conf}(p_j) = \{ p_k \mid p_k \perp p_j \}
\end{equation}
\textit{(iii) Negative Knowledge}: Identifies pass sequences that cause performance regressions or undesired effects indicated by evaluation function $\mathcal{E}$ and threshold $\epsilon$. These sequences are cataloged in $\mathcal{K}$ as \textit{non-optimal sequences}, and the knowledge base prevents their future use. We formalize this as:
\begin{equation}
    \mathcal{K}_{\text{neg}} = \{(p_1, p_2, \dots, p_k) \mid \mathcal{E}(p_1, p_2, \dots, p_k) < \epsilon\},
\end{equation}

\paragraph{Context-Aware Dataset Construction.} To facilitate agent training, we construct a high-quality reasoning dataset $\mathcal{D} = \{(\mathbf{x}_i, \mathcal{T}_i, \pi_i^*, \mathcal{E}_i)\}_{i=1}^N$. \textit{Program representation $\mathbf{x}_i$} captures the code’s statistical features via AutoPhase \cite{autophase}, ensuring it remains within the context window limit. \textit{Reasoning process $\mathcal{T}_i$} encapsulates prompt templates that guide the model through deep reasoning and iterative optimization, fostering contextual awareness. \textit{Pass sequence $\pi_i^*$} represents the optimal pass sequence derived from expert annotations or learned heuristics, offering guidance during training. \textit{Optimization effect $\mathcal{E}_i$} quantifies the code size reduction after applying $\pi_i^*$, reflecting the program’s efficiency improvement. This context-aware dataset enables the compiler agent to learn the relationships between code features, reasoning steps, pass trajectories, and optimization effects, supporting adaptive reasoning across diverse program context.

\begin{tcolorbox}[title={An Example of Training Sample},   
                  coltitle=white,
                  colback=blue!4,
                  colframe=blue!80!black,
                  fonttitle=\bfseries,
                  left=2mm,right=2mm,top=1mm,bottom=1mm,
                  boxrule=0.8pt]
\verb|"Program Representation"|: "NumEdges": 63, "numConstZeroes": 34, "UncondBranches": 35, "NumAShrInst": 0, ... \\ \\
\verb|"Reasoning Process"|: Analyzing the autophase features: Total instructions = 1200, Total blocks = 200, Memory instructions = 800, Branch count = 150. Given the high number of memory instructions and blocks, I will prioritize memory and control-flow optimizations. Let me first try an initial pass sequence and check its effectiveness...  \\  \\
\quad \verb|"Pass Sequence"|: ["--inferattrs", "--dse", "--mldst-motion", "--mergefunc", "--instsimplify", "--correlated-propagation", "--slp-vectorizer", "--early-cse-memssa", "--gvn", ...]  \\  \\
\quad \verb|"Optimization Effect"|: "Status": "success", "Improvement (over\_oz)": 17.72\%.  
\end{tcolorbox}

\subsection{Knowledge-driven Adaptive Pass Generation}  
AwareCompiler empowers agents with \textit{context awareness} through a knowledge-driven adaptive reasoning paradigm, enabling them to dynamically generate the optimization sequence based on code feature and knowledge assistance, thereby mitigating the brittleness of direct optimization.

\paragraph{Code Feature Extraction.} The first step involves identifying critical syntactic and semantic features that influence optimization decisions, such as instruction count, control flow complexity, and memory access patterns. Given a  source code \( x \in \mathcal{X} \), we define a feature extraction function \( \mathcal{F}(\cdot) \) that maps \( x\) to a higher-dimensional feature space \( \mathcal{Z} \) via a nonlinear transformation:
\begin{equation}
    \mathbf{z} = \mathcal{F}(x) \in \mathcal{Z}, \quad \mathcal{Z} = \left\{ z_1, z_2, \dots, z_m \right\}, \quad z_i = \phi_i(z), \quad \phi_i: \mathcal{Z} \to \mathbb{R}^{d_i}, \quad d_i > 1
\end{equation}
The feature vector \( \mathbf{z} \) combines high-dimensional statistical and latent semantic features, with each \( z_i \) representing an embedding from the feature map \( \phi_i \). The extracted \( \mathbf{z} \) provides a comprehensive representation of the program's structure, which serves as the foundation for the subsequent domain knowledge retrieval and optimization sequence generation.

\paragraph{Domain Knowledge Retrieval.} The second step in reasoning involves retrieving relevant knowledge from the compiler knowledge base \( \mathcal{K} \) using a rank-based fusion mechanism, $\text{RankScore}(\cdot)$, which balances (i) the similarity between the code's feature vector \( \mathbf{z} \) and the pass sequence \( \pi \), quantified by \( \sum_{z_i \in \mathbf{z}} \text{sim}(\phi_i(\mathbf{z}), \phi(\pi)) \), and (ii) the minimization of the code size \( \mathbb{E}_{\mathbf{z} \sim \mathcal{Z}}\left[ L_{\text{size}}(\mathbf{z}, \pi) \right] \), guiding the agent to select the most relevant knowledge for optimization:
\begin{equation}
    \mathcal{R}(\mathbf{z}, \mathcal{K}) = \text{Top-K} \left( \left\{ \text{RankScore} \left( \sum_{z_i \in \mathbf{z}, \pi \in \mathcal{K}} \text{sim}(\phi_i(\mathbf{z}), \phi(\pi)), \mathbb{E}_{\mathbf{z} \sim \mathcal{Z}}\left[ \mathcal{L}_{\text{size}}(\mathbf{z}, \pi) \right] \right) \right\} \right)
\end{equation}

\paragraph{Pass Sequence Generation.} The final step involves generating an optimal pass sequence \( \pi^* \) that minimizes the code size \( L_{\text{size}}(x, \pi) \), while satisfying the semantic constraints. Specifically, the indicator function \( \mathbb{I}[\cdot] \) ensures the validity of the pass sequence by respecting dependencies and avoiding conflicts. The whole process can be formulated as:
\begin{equation}
    \pi^* = \arg\min_{\pi \in \mathcal{R}(\mathbf{z}, \mathcal{K})} \mathbb{E}_{\pi} \left[ L_{\text{size}}(x, \pi) \right] \text{subject to} \sum_{p_i, p_j \in \pi} \mathbb{I}[p_i \in \text{deps}(p_j)] \quad \text{and} \quad \mathbb{I}[p_i \in \text{conf}(p_j)]
\end{equation}


\subsection{Data-driven Hybrid Training Pipeline}
To integrate knowledge and data effectively, AwareCompiler adopts a two-stage training approach. The first stage, Supervised Fine-Tuning (SFT), trains the model to follow specific reasoning format and identify effective pass sequences quickly. The second stage introduces Reinforcement Learning (RL), where a composite reward function balances performance gains with penalties for violations, evolving the agent from accuracy to effectiveness.

\paragraph{Supervised Fine-Tuning.} During the SFT phase, the agent learns optimization heuristics to solve various tasks. This enables the agent to internalize fundamental optimization patterns, follow specific reasoning formats, invoke knowledge base retrieval, and generate valid sequences. The SFT objective is to align the agent's policy \( \hat{\pi}^{SFT} \) with expert-defined strategies $\hat{\pi}^{SFT} = \arg\min_{\pi \in \mathcal{P}} \mathcal{L}_{size}(\pi, \mathbf{x}, \mathcal{D})$.



\paragraph{Reinforcement Learning.} The RL phase refines the agent's decision-making by exploring optimization paths and receiving rewards that balance performance with constraint penalties. The goal is to learn a policy \( \hat{\pi}^{RL} \) that maximizes long-term cumulative rewards:
\begin{equation}
     \hat{\pi}^{RL} = \arg\max_{\pi} \mathbb{E}_{\pi} \left[ \sum_{t=0}^{T} \gamma^t r_t \right]
\end{equation}
where \( r_t \) is the reward at time \( t \), and \( \gamma \) is the discount factor. The composite reward function aligns outputs with format standardization, answer validity, and performance improvement: 
\begin{itemize}[leftmargin=1.5em]
    \item \textit{Format Reward}: Encourage the model follows structured decision-making: reasoning within \texttt{<think>}, retrieve knowledge base within \texttt{<tool\_call>}, and generate pass sequence within \texttt{<answer>}. A score of 1 point is awarded for valid format and 0 otherwise. 
    \item \textit{Answer Reward}: Validates the answer through a compilation test, with schema constraints and tool protocols are checked respectively, ensuring the generated optimization passes are syntactically and structurally correct.
    \item \textit{Performance Reward}: Measures the reduction in code size $\Delta IC = \{IC_{\text{before}} - IC_{\text{after}}\} / {IC_{\text{before}}}$, as $IC_{\text{before}}, IC_{\text{after}}$ denote the instruction counts before and after applying the candidate sequence, respectively.
\end{itemize}

\paragraph{End-to-End Training and Synergy.} The two-stage, synergistic training process, consisting of SFT followed by RL, combines stable learning with dynamic adaptation, enabling AwareCompiler to progressively learn from simple heuristics to optimized solutions. The joint objective of the whole learning process can be described as:
\begin{equation}
    \pi^{\text{final}} = \arg\max_{\pi} \left( -\mathcal{L}_{size}(\pi, \mathbf{x}, \mathcal{D}_{SFT}) + \mathbb{E}_{\pi} \left[ \sum_{t=0}^{T} \gamma^t r_t \right] \right)
\end{equation}
This synergy ensures the agent satisfies constraints while significantly improving performance.

\begin{table*}[htbp]
\centering
\caption{Comparison of code size reduction across heuristic, ML-based, LLM-assisted, and our proposed AwareCompiler on various benchmarks. Results are reported as percentage reductions in LLVM IR count, where higher values indicate greater effectiveness.}
\label{tab:main}
\resizebox{\textwidth}{!}{
\begin{tabular}{l|ccccccc|c}
\toprule
\textbf{Method} & \textbf{blas} & \textbf{cbench} & \textbf{chstone} & \textbf{mibench} & \textbf{npb} & \textbf{opencv} & \textbf{tensorflow} & \textbf{Avg.} \\
\midrule
\multicolumn{9}{l}{\textit{Heuristic Optimization}} \\
Opt -O1 & 4.78\% & 50.87\% & 53.10\% & 59.05\% & 36.52\% & 3.11\% & 1.74\% & 29.88\% \\
Opt -O2 & 5.88\% & 27.23\% & 31.98\% & 44.39\% & 34.26\% & 2.00\% & 1.25\% & 21.00\% \\
Opt -O3 & 4.64\% & 15.96\% & 28.87\% & 21.46\% & 27.93\% & -8.90\% & -0.77\% & 12.74\% \\
Opt -Oz & 5.70\% & 51.93\% & 50.56\% & 58.29\% & 38.21\% & 3.38\% & 1.77\% & 29.98\% \\
\midrule
\multicolumn{9}{l}{\textit{LLM-assisted Optimization}} \\
GPT-5                   & 4.35\% & 25.69\% & 11.24\% & 27.18\% & 19.70\% & 0.34\%  & 1.90\%  & 12.91\% \\
Gemini-2.5-pro          & 1.21\% & 17.93\% & 25.14\% & 39.61\% & 35.91\% & -0.36\% & -0.26\% & 17.03\% \\
DeepSeek-V3             & 1.05\% & 49.05\% & 48.55\% & 57.62\% & 31.21\% & -1.57\% & -0.28\% & 26.52\% \\
Claude-opus-4           & 2.58\% & 47.44\% & 48.87\% & 58.65\% & 0.00\% & 1.51\%  & 0.25\%  & 22.76\% \\
GLM-4.5                 & 2.41\% & 39.79\% & 42.93\% & 48.72\% & 33.81\% & 1.05\%  & 0.20\%  & 24.13\% \\
Hunyuan-A13B-Instruct   & 0.67\% & 15.55\% & 28.93\% & 25.03\% & 14.78\% & 1.08\%  & 0.62\%  & 12.38\% \\
Kimi-Dev-72B            & 1.03\% & 46.87\% & 47.67\% & 54.71\% & 34.20\% & 2.87\%  & 1.39\%  & 26.96\% \\
Qwen3-235B-A22B         & 0.56\% & 30.60\% & 21.29\% & 19.23\% & 14.51\% & 0.99\%  & 0.78\%  & 12.57\% \\
Qwen3-Coder-480B-A35B   & 1.47\% & 37.04\% & 36.89\% & 43.16\% & 23.38\% & -3.78\% & 0.68\%  & 19.83\% \\
QwenLong-L1-32B         & 0.06\% & 13.04\% & 16.73\% & 16.73\% & 12.14\% & 0.60\%  & 0.17\%  & 8.50\% \\
\midrule
\multicolumn{9}{l}{\textit{Ours}} \\
\rowcolor{blue!5}  \textbf{AwareCompiler-1.5B}   & 5.45\% & 51.93\% & 49.91\% & 58.29\% & 38.73\% & 3.30\% & 2.60\% & 30.03\% \\
\rowcolor{blue!10} \textbf{AwareCompiler-3B}    & 6.32\% & 47.67\% & 45.22\% & 49.97\% & 40.09\% & 4.47\% & 1.29\% & 27.86\% \\
\rowcolor{blue!20} \textbf{AwareCompiler-7B}    & 4.94\% & 50.55\% & 49.44\% & 57.04\% & 39.36\% & 2.88\% & 2.06\% & 29.47\% \\
\bottomrule
\end{tabular}
}
\end{table*}

\section{Experiment}
We evaluate the effectiveness of \textbf{AwareCompiler} through a series of empirical studies aimed at addressing the following research questions (RQs): \textbf{RQ1:} How does AwareCompiler perform in comparison to existing baselines in compiler optimization? \textbf{RQ2:} Does knowledge-driven reasoning in AwareCompiler effectively address the semantic misalignment challenge? \textbf{RQ3:} What is the impact of the various driving methods (data and knowledge) employed in AwareCompiler?  \textbf{RQ4:} How do the different reward functions in AwareCompiler contribute to its optimization performance? \textbf{RQ5:} What are the key details of AwareCompiler’s reasoning and optimization process?

\subsection{Experimental Setup}
\paragraph{Benchmarks.} We evaluate AwareCompiler on a diverse collection of benchmark suites: blas \cite{blas}, npb \cite{npb}, cbench \cite{cbench}, opencv \cite{opencv}, mibench \cite{mibench}, chstone \cite{chstone}, tensorflow \cite{tensorflow}, providing a comprehensive evaluation across domains.

\paragraph{Baselines.} We compare AwareCompiler against several compiler optimization methods, including: (1) \textbf{Heuristic Methods}: expert optimization based on empirical knowledge, represented by \texttt{opt -O1}, \texttt{opt -O2}, \texttt{opt -O3}, and \texttt{opt -Oz} \cite{o3}; (2) \textbf{Vanilla LLMs}: mainstream LLMs using few-shot prompting with chain-of-thought, such as GPT-5 \cite{gpt5}, DeepSeek-V3 \cite{deepseek}, and Gemini-2.5 \cite{gemini}. Notably, we exclude other ML-based methods due to non-public datasets.

\paragraph{Training Details.} AwareCompiler was trained using the Qwen2.5-instruct models (1.5B, 3B, 7B) \cite{qwen2025qwen25technicalreport}. The training pipeline includes 2,000+ SFT samples for policy initialization, followed by RL on 15,000+ samples for policy enhancement. Experiments were conducted on Intel Xeon Gold 6430 servers (128 cores, 1TB RAM) with NVIDIA H100 GPUs (4×80GB HBM3).

\paragraph{Evaluation Details.} We use LLVM \cite{LLVM} IR count as the representation of code size, and average percentage of IR instruction reduction as the metric of optimization. All baseline models operate within a fixed optimization space consisting of 124 LLVM 10.0.0 optimization passes, ensuring a fair comparison of performance across different methods.

\subsection{Main Results (RQ1)}
\textbf{Experiment Objective.} This experiment evaluates AwareCompiler's effectiveness in reducing code size across various benchmarks, using LLVM IR count as the optimization metric, where higher values indicate better performance.

\textbf{Result Analysis.} As shown in Table~\ref{tab:main}, AwareCompiler consistently outperforms LLM-assisted methods, achieving reductions comparable to expert-level optimizations. Compared to advanced models like GPT-5 (12.91\%) and DeepSeek-V3 (26.52\%), AwareCompiler demonstrates significant improvements, even with smaller models. This highlights the effectiveness of its knowledge-data-driven approach, where knowledge-driven reasoning and data-driven optimization are seamlessly integrated. The key advantage of AwareCompiler lies in its ability to perform dynamic, context-aware optimizations, generating valid and effective pass sequences for complex tasks while reducing reliance on brute-force methods typically employed by LLMs.

\textbf{Summary.} AwareCompiler's ability to achieve expert-level optimization with its unique knowledge-data-driven framework across various benchmarks, not only positions it as an effective and reliable solution for compiler optimization, but also demonstrates the potential of integrating structured knowledge with learning-based systems to overcome the scalability challenges faced by traditional optimization methods.

\subsection{Semantic Misalignment Analysis (RQ2)}
\begin{figure*}[htbp]
  \centering
  \includegraphics[width=\linewidth]{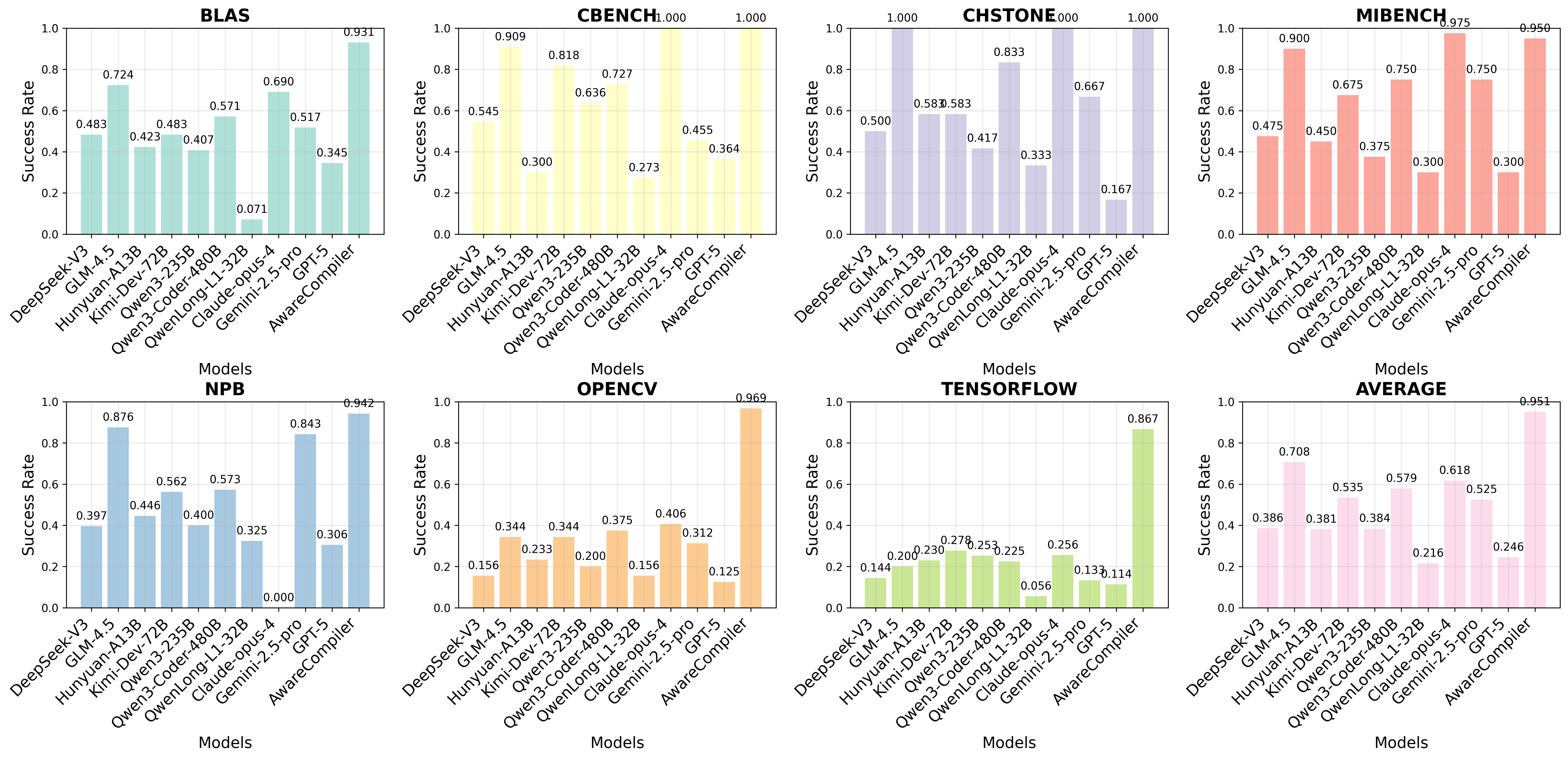}
  \caption{Comparison of success rates in generating valid pass sequences across different models.}
  \label{fig:success}
\end{figure*}

\textbf{Experiment Objective.} We evaluate the effectiveness of the AwareCompiler by measuring its success rate across various benchmarks. The success rate quantifies the proportion of valid optimization passes generated, which directly impacts the quality and efficiency of compiler optimization. 
 
\textbf{Result Analysis.} As shown in Figure~\ref{fig:success}, AwareCompiler achieves the highest success rates across benchmarks, with near-perfect results in \texttt{CBench} and \texttt{CHSTONE}. This highlights the effectiveness of its knowledge-driven adaptive pass generation in mitigating the \textbf{semantic misalignment challenge}. In contrast, LLMs like GPT-5, Gemini-2.5-pro, and Claude-opus-4 show lower success rates, especially in \texttt{OpenCV} and \texttt{TensorFlow}, reflecting their limitations in building accurate semantic links between program representations and optimization passes.

\textbf{Summary.} The superior success rates of AwareCompiler demonstrate the effectiveness of its integrated knowledge base and knowledge-driven adaptive pass generation mechanism, which bridges the semantic gap between abstract program representations and concrete optimization passes, significantly enhancing the efficiency, reliability, and robustness of compiler optimizations.

\subsection{Ablation Study (RQ3)}
\begin{table*}[htbp]
\centering
\caption{Ablation study for AwareCompiler, comparing optimization performance across various configurations: with full knowledge and data, and without knowledge, data, or both.}
\label{tab:ablation}
\resizebox{\textwidth}{!}{
\begin{tabular}{l|ccccccc|c}
\toprule
\textbf{Method} & \textbf{blas} & \textbf{cbench} & \textbf{chstone} & \textbf{mibench} & \textbf{npb} & \textbf{opencv} & \textbf{tensorflow} & \textbf{Avg.} \\
\midrule
\rowcolor{blue!5} \textbf{AwareCompiler-1.5B}   & 5.45\% & 51.93\% & 49.91\% & 58.29\% & 38.73\% & 3.30\% & 2.60\% & 30.03\% \\
w/o knowledge                                   & 5.32\% & 46.20\% & 46.60\% & 54.50\% & 34.1\% & 2.60\% & 1.35\% & 27.24\% \\
w/o data                                        & 4.87\% & 51.92\% & 50.56\% & 55.98\% & 37.77\% & 2.82\% & 1.69\% & 29.37\% \\
w/o knowledge \& data                           & 0.96\% & 36.68\% & 24.59\% & 33.68\% & 21.37\% & 0.67\% & 0.93\% & 16.98\% \\
\midrule
\rowcolor{blue!10} \textbf{AwareCompiler-3B}    & 6.32\% & 47.67\% & 45.22\% & 49.97\% & 40.09\% & 4.47\% & 1.29\% & 27.86\% \\
w/o knowledge                                   & 0.66\% & 38.60\% & 28.54\% & 35.77\% & 38.83\% & 3.18\% & 0.81\% & 20.91\% \\
w/o data                                        & 5.25\% & 46.95\% & 40.00\% & 48.92\% & 36.49\% & 2.47\% & 1.09\% & 25.88\% \\
w/o knowledge \& data                           & 0.21\% & 3.83\% & 16.95\% & 12.14\% & 5.09\% & -0.45\% & 0.14\% & 5.42\% \\
\midrule
\rowcolor{blue!20} \textbf{AwareCompiler-7B}    & 4.94\% & 50.55\% & 49.44\% & 57.04\% & 39.36\% & 2.88\% & 2.06\% & 29.47\% \\
w/o knowledge                                   & 1.79\% & 50.43\% & 48.73\% & 54.78\% & 40.10\% & 2.39\% & 1.24\% & 28.49\% \\
w/o data                                        & 5.42\% & 50.43\% & 46.34\% & 57.01\% & 38.01\% & 3.03\% & 1.68\% & 28.82\% \\
w/o knowledge \& data                           & 0.96\% & 36.58\% & 24.59\% & 33.68\% & 21.37\% & 0.67\% & 0.93\% & 16.97\% \\
\bottomrule
\end{tabular}
}
\end{table*}

\textbf{Experiment Objective.} We conducted an ablation study to assess the contributions of key components in AwareCompiler: the knowledge-driven adaptive reasoning ("w/o knowledge" by removing the knowledge base), the data-driven training pipeline ("w/o data" by using only SFT), and their combined effect ("w/o knowledge \& data" by using the base model).

\textbf{Result Analysis.} As shown in Table \ref{tab:ablation}, "w/o knowledge" leads to a noticeable performance drop, emphasizing the importance of context-aware knowledge retrieval. The "w/o data" exhibits a similar decrease, highlighting the value of the RL training. The "w/o knowledge \& data" shows the largest reduction, underscoring the necessity of both components for effective optimization. 

\textbf{Summary.} These results confirm that AwareCompiler’s synergistic knowledge-data-driven method, which combines knowledge-driven reasoning with data-driven learning, effectively overcomes the \textbf{ineffective interaction challenge}, resulting in superior pass sequences with improved quality.

\subsection{Reward Sparsity Analysis (RQ4)}
\begin{figure}[htbp]
    \footnotesize
    \centering
    \vspace{-6mm}
    \subfloat[Impact of reward schemes on optimization.]{
        \begin{minipage}{0.45\textwidth}
            \vspace{-35mm}
            \centering
            \renewcommand\arraystretch{1.15}
            \setlength{\tabcolsep}{3mm}{
                \begin{tabular}{cc}
                \toprule
                \textbf{Reward Scheme} & \textbf{Performance Gain} \\
                \midrule
                w/o $R_{format}$ & 8.02\%  \\
                w/o $R_{performance}$ & 12.50\%  \\
                w/o $R_{answer}$ & 18.10\%  \\
                \textbf{Full Reward} & 30.03\% \\
                \bottomrule
                \end{tabular}
            }
        \end{minipage}
    }
    \hspace{1mm}
    \subfloat[Impact of reward schemes on training.]{
        \includegraphics[width=0.45\textwidth]{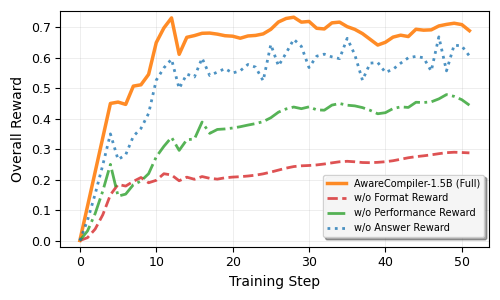}
    }
    \caption{Impact of reward functions on optimization performance and training efficiency.}
    \label{fig:rq4}
\end{figure}

\textbf{Experiment Objective.} This experiment evaluates the effectiveness of AwareCompiler in addressing the reward sparsity challenge by examining the impact of different reward function configurations on optimization performance and training efficiency.

\textbf{Result Analysis.} As shown in Figure~\ref{fig:rq4}, the full reward configuration achieves a performance gain of 30.03\%. Removing \( R_{\text{format}} \), \( R_{\text{answer}} \), or \( R_{\text{performance}} \) results in performance degradation. Furthermore, the full reward configuration leads to faster model convergence and a higher cumulative reward.

\textbf{Summary.} AwareCompiler effectively addresses the \textbf{reward sparsity challenge} by utilizing a comprehensive reward function that balances format standardization, answer validity, and performance improvement, thereby ensuring both effective optimizations and efficient learning.

\subsection{Case Study: Detailed Demonstration of AwareCompiler (RQ5)}
\begin{tcolorbox}[
  colback=cyan!5!white,     
  colframe=teal!90!black,   
  coltitle=white,           
  title=A case study of AwareCompiler's agentic workflow,  
  fonttitle=\bfseries,      
  sharp corners=south,      
  boxrule=1pt               
]
\textcolor{blue}{\textless think\textgreater}: "Based on autophase features, I generate an initial optimization sequence to reduce the program's instruction count. I will begin with a common heuristic approach." \\
\textcolor{magenta}{\textless tool\_call\textgreater}: "Let me verify the performance of the initial heuristic sequence by using the `instrcount` tool to compare it against the baseline optimization (`-Oz`)." \\
\textcolor{blue}{\textless think\textgreater}: "The initial optimization attempt has resulted in no performance improvement, indicating the need for a more tailored approach." \\
\textcolor{blue}{\textless think\textgreater}: "Both heuristic attempts have failed. It's clear that the generalized strategies are insufficient for this particular program. I will query the knowledge base for expert advice to generate a more effective pass sequence." \\
\textcolor{magenta}{\textless tool\_response\textgreater}: "The knowledge base suggests using the '--gvn' pass to eliminate redundant computations, which has been empirically validated for similar programs." \\
\textcolor{blue}{\textless think\textgreater}: "I'll integrate the recommended pass into the sequence. This sequence should meet the optimization goal based on the program's characteristics." \\
\textcolor{brown}{\textless answer\textgreater}: ["--loop-sink", "--gvn", "--ipsccp", ...] \\
\end{tcolorbox}

\textbf{Experiment Objective.} This case study evaluates AwareCompiler’s ability to improve performance when heuristics fail and whether the knowledge base can guide the agent to achieve better results.

\textbf{Result Analysis.} The agent initially generated an optimization sequence based on autophase features with no improvement. After consulting the knowledge base, the agent applied a tailored sequence, resulting in a 3.2\% improvement, highlighting the effectiveness of the knowledge base.

\textbf{Summary.} This case study underscores how knowledge-driven adaptive pass generation enables AwareCompiler to overcome the limitations of heuristics and achieve consistent performance gains.

\section{Related Work}
\textbf{Compiler Optimization.} Compiler optimization can be classified into three categories: Heuristic optimizations rely on predefined rules and search-based methods to enhance performance \cite{RIO, GA}. Machine learning-based methods use models trained on domain-specific datasets to generate optimization sequences, but face scalability challenges in larger, more diverse codebases \cite{autophase, BOCA, Comptuner, CompilerDream, pan2025towards}. LLM-assisted methods enable agents to autonomously explore optimizations via specialized training pipelines \cite{cummins2023large, gong2025languagemodelscodeoptimization}. AwareCompiler introduces a synergistic knowledge-data-driven approach that dynamically generates optimization passes, overcoming the limitations of rigid rule-based or static learning methods.

\textbf{Agentic Language Models.} Recent advancements in LLMs have led to agentic language models capable of autonomous reasoning, planning, and tool interaction \cite{huang2024understanding,zhao2024expel}. These language agents generate hypotheses and interact with external environments to achieve goals \cite{mei2024aios,hong2024data}. A key challenge is grounding their reasoning in specific domains to ensure effective and valid actions \cite{shang2024agentsquare, Compiler-R1}. AwareCompiler applies this agentic paradigm to compiler optimization, navigating the structured, vast optimization space.

\section{Conclusion}
We introduce \textbf{AwareCompiler}, an agentic framework for compiler optimization that leverages a synergistic knowledge-data-driven approach. By integrating knowledge-driven reasoning with data-driven learning, AwareCompiler effectively addresses challenges such as semantic misalignment, inefficient agent-environment interactions, and reward sparsity. Extensive experiments across multiple benchmarks demonstrate that AwareCompiler outperforms existing methods, achieving significant code size reductions while ensuring the validity of optimizations. AwareCompiler lays a robust foundation for more efficient, flexible, and automated compiler optimization in the future.

\section*{ETHICS STATEMENT}
AwareCompiler leverages LLMs and domain-specific knowledge and data to automate compiler optimization. While this framework enhances efficiency, ethical concerns must be considered. In particular, the potential for producing suboptimal or invalid optimization passes could destabilize system performance. AwareCompiler mitigates this risk through its knowledge-driven reasoning mechanism and continuous validation processes, but ongoing monitoring and verification are necessary. The framework’s reliance on large-scale datasets and LLMs also raises questions regarding data quality and model biases, which must be addressed through responsible usage, transparent methodologies, and regular updates to the underlying models. Ethical deployment requires careful attention to the accuracy, reliability, and potential environmental impact of optimizations generated by AwareCompiler.

\section*{REPRODUCIBILITY}
To ensure the reproducibility of AwareCompiler, we provide full access to the source code and experimental setup. All models, training data, and scripts are publicly available at the following URL: https://github.com/LHY-24/AwareCompiler. This repository includes detailed instructions for setting up the required environment, performing the experiments, and replicating the results across standard benchmarks. By sharing these resources, we aim to facilitate the verification and replication of our results, supporting further advancements in the field of compiler optimization.

\bibliography{iclr2026_conference}

\begin{thebibliography}{33}
\providecommand{\natexlab}[1]{#1}
\providecommand{\url}[1]{\texttt{#1}}
\expandafter\ifx\csname urlstyle\endcsname\relax
  \providecommand{\doi}[1]{doi: #1}\else
  \providecommand{\doi}{doi: \begingroup \urlstyle{rm}\Url}\fi

\bibitem[Abadi et~al.(2016)Abadi, Agarwal, Barham, Brevdo, Chen, Chen, Corrado,
  Davis, Dean, Gardner, Harris, Goodfellow, Irving, Isard, Jia, Jozefowicz,
  Kaiser, Kudlur, Levenberg, Mané, Monga, Moore, Murray, Olah, Schuster,
  Vishwanath, Zhang, and Zhou]{tensorflow}
Martín Abadi, Ashish Agarwal, Paul Barham, Eugene Brevdo, Zhifeng Chen,
  Chengyue Chen, Geoffrey~S. Corrado, Andy Davis, Jeffrey Dean, Mathews
  Gardner, Greg Harris, Ian~J. Goodfellow, Geoffrey Irving, Michael Isard,
  Yangqing Jia, Rafal Jozefowicz, Lukasz Kaiser, Manjunath Kudlur, Josh
  Levenberg, Dandelion Mané, Rajat Monga, Sherry Moore, Derek Murray, Chris
  Olah, Mike Schuster, Rajiv Vishwanath, Shuzhao Zhang, and Yuan Zhou.
\newblock Tensorflow: Large-scale machine learning on heterogeneous distributed
  systems.
\newblock In \emph{Proceedings of the 12th USENIX Symposium on Operating
  Systems Design and Implementation (OSDI)}, pp.\  265--283. USENIX
  Association, 2016.
\newblock URL
  \url{https://www.usenix.org/conference/osdi16/technical-sessions/presentation/abadi}.
\newblock https://www.tensorflow.org/.

\bibitem[Aho et~al.(2006)Aho, Lam, Sethi, and Ullman]{aho2006compilers}
Alfred~V. Aho, Monica~S. Lam, Ravi Sethi, and Jeffrey~D. Ullman.
\newblock \emph{Compilers: Principles, Techniques, and Tools (2nd Edition)}.
\newblock Addison-Wesley Longman Publishing Co., Inc., USA, 2006.
\newblock ISBN 0321486811.

\bibitem[Ansel et~al.(2014)Ansel, Kamil, Veeramachaneni, Ragan-Kelley, Bosboom,
  O'Reilly, and Amarasinghe]{opentuner}
Jason Ansel, Shoaib Kamil, Kalyan Veeramachaneni, Jonathan Ragan-Kelley,
  Jeffrey Bosboom, Una-May O'Reilly, and Saman Amarasinghe.
\newblock Opentuner: An extensible framework for program autotuning.
\newblock In \emph{Proceedings of the 23rd international conference on Parallel
  architectures and compilation}, pp.\  303--316, 2014.

\bibitem[Bailey et~al.(1995)Bailey, Barszcz, Magro, Kenhro, and Snavely]{npb}
David~H. Bailey, Eric Barszcz, Thomas~E. Magro, Kenhro, and Alan~S. Snavely.
\newblock The nas parallel benchmarks.
\newblock Technical Report RNR-91-004, NASA Ames Research Center, 1995.
\newblock URL \url{https://www.nas.nasa.gov/publications/npb.html}.

\bibitem[Bergstra et~al.(2011)Bergstra, Bardenet, Bengio, and K{\'e}gl]{tpe}
James Bergstra, R{\'e}mi Bardenet, Yoshua Bengio, and Bal{\'a}zs K{\'e}gl.
\newblock Algorithms for hyper-parameter optimization.
\newblock \emph{Advances in neural information processing systems}, 24, 2011.

\bibitem[Bradski(2000)]{opencv}
Gary Bradski.
\newblock The opencv library.
\newblock In \emph{Intel Developer Forum}, 2000.
\newblock URL \url{https://opencv.org/}.

\bibitem[Chen et~al.(2021)Chen, Xu, Chen, and Zhang]{BOCA}
Junjie Chen, Ningxin Xu, Peiqi Chen, and Hongyu Zhang.
\newblock Efficient compiler autotuning via bayesian optimization.
\newblock In \emph{2021 IEEE/ACM 43rd International Conference on Software
  Engineering (ICSE)}, pp.\  1198--1209. IEEE, 2021.

\bibitem[Chen \& Karp(1997)Chen and Karp]{cbench}
Liangyong Chen and Robert~M. Karp.
\newblock Cbench: A benchmarking suite for evaluating compiler optimization.
\newblock In \emph{Proceedings of the ACM SIGPLAN 1997 Conference on
  Programming Language Design and Implementation (PLDI)}, pp.\  40--49. ACM,
  1997.
\newblock \doi{10.1145/258915.258920}.
\newblock URL \url{https://dl.acm.org/doi/10.1145/258915.258920}.

\bibitem[Chen et~al.(2012)Chen, Fang, Huang, Eeckhout, Fursin, Temam, and
  Wu]{RIO}
Yang Chen, Shuangde Fang, Yuanjie Huang, Lieven Eeckhout, Grigori Fursin,
  Olivier Temam, and Chengyong Wu.
\newblock Deconstructing iterative optimization.
\newblock \emph{ACM Transactions on Architecture and Code Optimization (TACO)},
  9\penalty0 (3):\penalty0 1--30, 2012.

\bibitem[Cummins et~al.(2021)Cummins, Wasti, Guo, Cui, Ansel, Gomez, Jain, Liu,
  Teytaud, Steiner, Tian, and Leather]{CompilerGym}
Chris Cummins, Bram Wasti, Jiadong Guo, Brandon Cui, Jason Ansel, Sahir Gomez,
  Somya Jain, Jia Liu, Olivier Teytaud, Benoit Steiner, Yuandong Tian, and Hugh
  Leather.
\newblock Compilergym: Robust, performant compiler optimization environments
  for ai research, 2021.
\newblock URL \url{https://arxiv.org/abs/2109.08267}.

\bibitem[Cummins et~al.(2023)Cummins, Seeker, Grubisic, Elhoushi, Liang,
  Roziere, Gehring, Gloeckle, Hazelwood, Synnaeve, et~al.]{cummins2023large}
Chris Cummins, Volker Seeker, Dejan Grubisic, Mostafa Elhoushi, Youwei Liang,
  Baptiste Roziere, Jonas Gehring, Fabian Gloeckle, Kim Hazelwood, Gabriel
  Synnaeve, et~al.
\newblock Large language models for compiler optimization.
\newblock \emph{arXiv preprint arXiv:2309.07062}, 2023.

\bibitem[Cummins et~al.(2024)Cummins, Seeker, Grubisic, Roziere, Gehring,
  Synnaeve, and Leather]{cummins2024meta}
Chris Cummins, Volker Seeker, Dejan Grubisic, Baptiste Roziere, Jonas Gehring,
  Gabriel Synnaeve, and Hugh Leather.
\newblock Meta large language model compiler: Foundation models of compiler
  optimization.
\newblock \emph{arXiv preprint arXiv:2407.02524}, 2024.

\bibitem[Deng et~al.(2025)Deng, Wu, Feng, Wang, and Long]{CompilerDream}
Chaoyi Deng, Jialong Wu, Ningya Feng, Jianmin Wang, and Mingsheng Long.
\newblock Compilerdream: Learning a compiler world model for general code
  optimization, 2025.
\newblock URL \url{https://arxiv.org/abs/2404.16077}.

\bibitem[Dongarra et~al.(1990)Dongarra, Duff, Hammarling, and Hanson]{blas}
Jack~J. Dongarra, Iain~S. Duff, Sven Hammarling, and Richard~A. Hanson.
\newblock A set of level 3 basic linear algebra subprograms.
\newblock In \emph{Proceedings of the ACM/IEEE Conference on Supercomputing},
  pp.\  1--7. ACM, 1990.
\newblock \doi{10.1109/SC.1990.125077}.
\newblock URL \url{https://ieeexplore.ieee.org/document/125077}.

\bibitem[et~al.(2025{\natexlab{a}})]{deepseek}
DeepSeek-AI et~al.
\newblock Deepseek-v3 technical report, 2025{\natexlab{a}}.
\newblock URL \url{https://arxiv.org/abs/2412.19437}.

\bibitem[et~al.(2025{\natexlab{b}})]{gemini}
Gheorghe~Comanici et~al.
\newblock Gemini 2.5: Pushing the frontier with advanced reasoning,
  multimodality, long context, and next generation agentic capabilities,
  2025{\natexlab{b}}.
\newblock URL \url{https://arxiv.org/abs/2507.06261}.

\bibitem[Garciarena \& Santana(2016)Garciarena and Santana]{GA}
Unai Garciarena and Roberto Santana.
\newblock Evolutionary optimization of compiler flag selection by learning and
  exploiting flags interactions.
\newblock In \emph{Proceedings of the 2016 on Genetic and Evolutionary
  Computation Conference Companion}, pp.\  1159--1166, 2016.

\bibitem[Gong et~al.(2025)Gong, Voskanyan, Brookes, Wu, Jie, Xu, Giavrimis,
  Basios, Kanthan, and Wang]{gong2025languagemodelscodeoptimization}
Jingzhi Gong, Vardan Voskanyan, Paul Brookes, Fan Wu, Wei Jie, Jie Xu, Rafail
  Giavrimis, Mike Basios, Leslie Kanthan, and Zheng Wang.
\newblock Language models for code optimization: Survey, challenges and future
  directions, 2025.
\newblock URL \url{https://arxiv.org/abs/2501.01277}.

\bibitem[Guthaus et~al.(2001)Guthaus, Baer, El-Ghazawi, Patel, and
  Schrag]{mibench}
David Guthaus, Jean-Loup Baer, Tarek S.~S. El-Ghazawi, Sanjay Patel, and Brian
  Schrag.
\newblock Mibench: A free, commercially representative embedded benchmark
  suite.
\newblock In \emph{Proceedings of the IEEE International Workshop on Workload
  Characterization (IWWC)}, pp.\  3--14. IEEE, 2001.
\newblock \doi{10.1109/IWWC.2001.948431}.
\newblock URL \url{https://ieeexplore.ieee.org/document/948431}.

\bibitem[Haj-Ali et~al.(2020)Haj-Ali, Huang, Xiang, Moses, Asanovic, Wawrzynek,
  and Stoica]{autophase}
Ameer Haj-Ali, Qijing~Jenny Huang, John Xiang, William Moses, Krste Asanovic,
  John Wawrzynek, and Ion Stoica.
\newblock Autophase: Juggling hls phase orderings in random forests with deep
  reinforcement learning.
\newblock \emph{Proceedings of Machine Learning and Systems}, 2:\penalty0
  70--81, 2020.

\bibitem[Hong et~al.(2024)Hong, Lin, Liu, Liu, Wu, Zhang, Wei, Li, Chen, Zhang,
  et~al.]{hong2024data}
Sirui Hong, Yizhang Lin, Bang Liu, Bangbang Liu, Binhao Wu, Ceyao Zhang,
  Chenxing Wei, Danyang Li, Jiaqi Chen, Jiayi Zhang, et~al.
\newblock Data interpreter: An llm agent for data science.
\newblock \emph{arXiv preprint arXiv:2402.18679}, 2024.

\bibitem[Huang et~al.(2024)Huang, Liu, Chen, Wang, Wang, Lian, Wang, Tang, and
  Chen]{huang2024understanding}
Xu~Huang, Weiwen Liu, Xiaolong Chen, Xingmei Wang, Hao Wang, Defu Lian, Yasheng
  Wang, Ruiming Tang, and Enhong Chen.
\newblock Understanding the planning of llm agents: A survey.
\newblock \emph{arXiv preprint arXiv:2402.02716}, 2024.

\bibitem[Krentel \& Ewing(1990)Krentel and Ewing]{o3}
Mark Krentel and Barry A.~P. Ewing.
\newblock The opt-o3 optimization framework for high-performance compilers.
\newblock In \emph{Proceedings of the International Conference on
  Supercomputing}, pp.\  123--134. ACM, 1990.
\newblock \doi{10.1145/99163.99175}.
\newblock URL \url{https://dl.acm.org/doi/10.1145/99163.99175}.

\bibitem[Lattner \& Adve(2004)Lattner and Adve]{LLVM}
C.~Lattner and V.~Adve.
\newblock Llvm: a compilation framework for lifelong program analysis \&
  transformation.
\newblock In \emph{International Symposium on Code Generation and Optimization,
  2004. CGO 2004.}, pp.\  75--86, 2004.
\newblock \doi{10.1109/CGO.2004.1281665}.

\bibitem[Mei et~al.(2024)Mei, Zhu, Xu, Hua, Jin, Li, Xu, Ye, Ge, and
  Zhang]{mei2024aios}
Kai Mei, Xi~Zhu, Wujiang Xu, Wenyue Hua, Mingyu Jin, Zelong Li, Shuyuan Xu,
  Ruosong Ye, Yingqiang Ge, and Yongfeng Zhang.
\newblock Aios: Llm agent operating system.
\newblock \emph{arXiv preprint arXiv:2403.16971}, 2024.

\bibitem[Miyamoto et~al.(2010)Miyamoto, Kondo, and Fukui]{chstone}
Hiroshi Miyamoto, Takanobu Kondo, and Kenjiro Fukui.
\newblock Chstone: A benchmark for high-level synthesis.
\newblock In \emph{Proceedings of the International Conference on
  Field-Programmable Logic and Applications (FPL)}, pp.\  388--393. IEEE, 2010.
\newblock \doi{10.1109/FPL.2010.141}.
\newblock URL \url{https://ieeexplore.ieee.org/document/5545432}.

\bibitem[OpenAI(2025)]{gpt5}
OpenAI.
\newblock Introducing gpt-5, 2025.
\newblock URL \url{https://openai.com/zh-Hans-CN/index/introducing-gpt-5/}.

\bibitem[Pan et~al.(2025{\natexlab{a}})Pan, Lin, Luo, Liu, Yao, Zhang, Xing,
  and Wu]{Compiler-R1}
Haolin Pan, Hongyu Lin, Haoran Luo, Yang Liu, Kaichun Yao, Libo Zhang, Mingjie
  Xing, and Yanjun Wu.
\newblock Compiler-r1: Towards agentic compiler auto-tuning with reinforcement
  learning, 2025{\natexlab{a}}.
\newblock URL \url{https://arxiv.org/abs/2506.15701}.

\bibitem[Pan et~al.(2025{\natexlab{b}})Pan, Wei, Xing, Wu, and
  Zhao]{pan2025towards}
Haolin Pan, Yuanyu Wei, Mingjie Xing, Yanjun Wu, and Chen Zhao.
\newblock Towards efficient compiler auto-tuning: Leveraging synergistic search
  spaces.
\newblock In \emph{Proceedings of the 23rd ACM/IEEE International Symposium on
  Code Generation and Optimization}, pp.\  614--627, 2025{\natexlab{b}}.

\bibitem[Qwen et~al.(2025)Qwen, :, Yang, Yang, Zhang, Hui, Zheng, Yu, Li, Liu,
  Huang, Wei, Lin, Yang, Tu, Zhang, Yang, Yang, Zhou, Lin, Dang, Lu, Bao, Yang,
  Yu, Li, Xue, Zhang, Zhu, Men, Lin, Li, Tang, Xia, Ren, Ren, Fan, Su, Zhang,
  Wan, Liu, Cui, Zhang, and Qiu]{qwen2025qwen25technicalreport}
Qwen, :, An~Yang, Baosong Yang, Beichen Zhang, Binyuan Hui, Bo~Zheng, Bowen Yu,
  Chengyuan Li, Dayiheng Liu, Fei Huang, Haoran Wei, Huan Lin, Jian Yang,
  Jianhong Tu, Jianwei Zhang, Jianxin Yang, Jiaxi Yang, Jingren Zhou, Junyang
  Lin, Kai Dang, Keming Lu, Keqin Bao, Kexin Yang, Le~Yu, Mei Li, Mingfeng Xue,
  Pei Zhang, Qin Zhu, Rui Men, Runji Lin, Tianhao Li, Tianyi Tang, Tingyu Xia,
  Xingzhang Ren, Xuancheng Ren, Yang Fan, Yang Su, Yichang Zhang, Yu~Wan,
  Yuqiong Liu, Zeyu Cui, Zhenru Zhang, and Zihan Qiu.
\newblock Qwen2.5 technical report, 2025.
\newblock URL \url{https://arxiv.org/abs/2412.15115}.

\bibitem[Shang et~al.(2024)Shang, Li, Zhao, Ma, Liu, Xu, and
  Li]{shang2024agentsquare}
Yu~Shang, Yu~Li, Keyu Zhao, Likai Ma, Jiahe Liu, Fengli Xu, and Yong Li.
\newblock Agentsquare: Automatic llm agent search in modular design space.
\newblock \emph{arXiv preprint arXiv:2410.06153}, 2024.

\bibitem[Zhao et~al.(2024)Zhao, Huang, Xu, Lin, Liu, and Huang]{zhao2024expel}
Andrew Zhao, Daniel Huang, Quentin Xu, Matthieu Lin, Yong-Jin Liu, and Gao
  Huang.
\newblock Expel: Llm agents are experiential learners.
\newblock In \emph{Proceedings of the AAAI Conference on Artificial
  Intelligence}, volume~38, pp.\  19632--19642, 2024.

\bibitem[Zhu et~al.(2024)Zhu, Hao, and Chen]{Comptuner}
Mingxuan Zhu, Dan Hao, and Junjie Chen.
\newblock Compiler autotuning through multiple-phase learning.
\newblock \emph{ACM Transactions on Software Engineering and Methodology},
  33\penalty0 (4):\penalty0 1--38, 2024.

\end{thebibliography}
\bibliographystyle{iclr2026_conference}

\newpage
\section*{Appendix}

\appendix

\section{Prompt Templates} \label{app:rawir-prompt}
\begin{lstlisting}
Act as a compiler optimization expert to find an optimal LLVM pass sequence that minimizes total instruction count.

LLVM IR features:
{formatted_features}
Initial instruction count: {TotalInsts}

Workflow:
1. <think>: Analyze features and reason about effective passes.
2. <tool_call>: Query lightrag_compiler_optimization for a recommended sequence.
3. <tool_call>: Verify the sequence using instrcount with {program_id}.
4. Interpret instrcount result: positive improvement_over_oz means the optimization is successful, negative means it is not. Adjust your strategy accordingly.
5. Output only your final pass sequence in <answer> tags as a JSON list.

Example output format: 
<|im_start|>assistant 
<think> Analyzing the autophase features, I notice a high number of memory instructions and branches. I will prioritize memory and control-flow optimizations. 
</think> 
<tool_call> 
{{"name": "lightrag_compiler_optimization", "arguments": {{"query": "{formatted_features}"}}}} 
</tool_call> 
<|im_end|> 

<|im_start|>user <tool_response> {{"recommended_pass_sequence": ["--inferattrs", "--dse", "--mldst-motion", "--mergefunc", ...], "performance_improvement": 0.42}} 
</tool_response> 
<|im_end|> 

<|im_start|>assistant 
<think> I will verify the recommended sequence using the instrcount tool. 
</think> 
<tool_call> 
{{"name": "instrcount", "arguments": {{"filename": "{program_id}", "optimization_flags": ["--inferattrs", "--dse", "--mldst-motion", "--mergefunc", ...]}}}} </tool_call> 
<|im_end|> 

<|im_start|>user 
<tool_response> {{"status": "success", "improvement_over_oz": 0.42}} 
</tool_response> 
<|im_end|> 

<|im_start|>assistant 
<answer>
["--inferattrs", "--dse", "--mldst-motion", "--mergefunc", ...]
</answer> 
<|im_end|>
\end{lstlisting}

\section{AutoPhase Feature Set}
\label{appendix:autophase_features}

As referenced in \textbf{Feature Extraction and Representation}, our framework utilizes the 56 statistical features from AutoPhase \cite{autophase} to represent programs compactly for the LLM. These features capture various aspects of the program's static structure and instruction mix. Table~\ref{tab:autophase_feature_list} provides a complete list of these features.

\begin{small} 
\begin{longtable}{clcl}
\caption{List of 56 AutoPhase Features Utilized.} \label{tab:autophase_feature_list} \\
\toprule
\textbf{Index} & \textbf{Feature Description} & \textbf{Index} & \textbf{Feature Description} \\
\midrule
\endfirsthead

\multicolumn{4}{c}%
{{\tablename\ \thetable{} -- continued from previous page}} \\
\toprule
\textbf{Index} & \textbf{Feature Description} & \textbf{Index} & \textbf{Feature Description} \\
\midrule
\endhead

\bottomrule
\multicolumn{4}{r}{{Continued on next page}} \\
\endfoot

\bottomrule
\endlastfoot

0 & BBs: total phi args \textgreater 5 & 28 & Number of And insts \\
1 & BBs: total phi args in [1,5] & 29 & BBs: instruction count in [15,500] \\
2 & BBs: count with 1 predecessor & 30 & BBs: instruction count \textless 15 \\
3 & BBs: count with 1 predecessor and 1 successor & 31 & Number of BitCast insts \\
4 & BBs: count with 1 predecessor and 2 successors & 32 & Number of Br insts \\
5 & BBs: count with 1 successor & 33 & Number of Call insts \\
6 & BBs: count with 2 predecessors & 34 & Number of GetElementPtr insts \\
7 & BBs: count with 2 predecessors and 1 successor & 35 & Number of ICmp insts \\
8 & BBs: count with 2 predecessors and successors & 36 & Number of LShr insts \\
9 & BBs: count with 2 successors & 37 & Number of Load insts \\
10 & BBs: count with \textgreater2 predecessors & 38 & Number of Mul insts \\
11 & BBs: Phi node count in range (0,3] per BB & 39 & Number of Or insts \\
12 & BBs: count with more than 3 Phi nodes & 40 & Number of PHI insts \\
13 & BBs: count with no Phi nodes & 41 & Number of Ret insts \\
14 & Number of Phi-nodes at beginning of BB & 42 & Number of SExt insts \\
15 & Number of branches & 43 & Number of Select insts \\
16 & Number of calls that return an int & 44 & Number of Shl insts \\
17 & Number of critical edges & 45 & Number of Store insts \\
18 & Number of edges & 46 & Number of Sub insts \\
19 & Occurrences of 32-bit integer constants & 47 & Number of Trunc insts \\ 
20 & Occurrences of 64-bit integer constants & 48 & Number of Xor insts \\ 
21 & Occurrences of constant 0 & 49 & Number of ZExt insts \\ 
22 & Occurrences of constant 1 & 50 & Number of basic blocks \\ 
23 & Number of unconditional branches & 51 & Number of instructions (all types) \\
24 & Binary operations with a constant operand & 52 & Number of memory instructions \\ 
25 & Number of AShr insts & 53 & Number of non-external functions \\
26 & Number of Add insts & 54 & Total arguments to Phi nodes \\
27 & Number of Alloca insts & 55 & Number of Unary operations \\
\end{longtable}
\end{small}

\newpage

\section{LLVM Optimization Passes}
\label{appendix:llvm_passes}

Our AwareCompiler framework and baseline models operate within an action space comprising 124 individual LLVM 10.0.0 \texttt{opt} transformation passes, resulting in a total of 125 distinct actions available to the optimization agent. Table~\ref{tab:llvm_pass_list} enumerates these passes and the \texttt{-Oz} along with their corresponding indices used within our system. These flags can typically be obtained from the CompilerGym LLVM environment.

{\fontsize{8}{9}\selectfont
\begin{longtable}{clclcl}
\caption{List of Utilized LLVM Compiler Passes and \texttt{-Oz} with Corresponding Indices.} \label{tab:llvm_pass_list} \\
\toprule
\textbf{Index} & \textbf{Flag} & \textbf{Index} & \textbf{Flag} & \textbf{Index} & \textbf{Flag} \\
\midrule
\endfirsthead 
\multicolumn{6}{c}%
{{\tablename\ \thetable{} -- continued from previous page}} \\
\toprule
\textbf{Index} & \textbf{Flag} & \textbf{Index} & \textbf{Flag} & \textbf{Index} & \textbf{Flag} \\
\midrule
\endhead 

\bottomrule
\multicolumn{6}{r}{{Continued on next page}} \\
\endfoot 

\bottomrule
\endlastfoot 

0 & add-discriminators & 42 & globalsplit & 84 & lower-expect \\
1 & adce & 43 & guard-widening & 85 & lower-guard-intrinsic \\
2 & aggressive-instcombine & 44 & hotcoldsplit & 86 & lowerinvoke \\
3 & alignment-from-assumptions & 45 & ipconstprop & 87 & lower-matrix-intrinsics \\
4 & always-inline & 46 & ipsccp & 88 & lowerswitch \\
5 & argpromotion & 47 & indvars & 89 & lower-widenable-condition \\
6 & attributor & 48 & irce & 90 & memcpyopt \\
7 & barrier & 49 & infer-address-spaces & 91 & mergefunc \\
8 & bdce & 50 & inferattrs & 92 & mergeicmps \\
9 & break-crit-edges & 51 & inject-tli-mappings & 93 & mldst-motion \\
10 & simplifycfg & 52 & instsimplify & 94 & sancov \\
11 & callsite-splitting & 53 & instcombine & 95 & name-anon-globals \\
12 & called-value-propagation & 54 & instnamer & 96 & nary-reassociate \\
13 & canonicalize-aliases & 55 & jump-threading & 97 & newgvn \\
14 & consthoist & 56 & lcssa & 98 & pgo-memop-opt \\
15 & constmerge & 57 & licm & 99 & partial-inliner \\
16 & constprop & 58 & libcalls-shrinkwrap & 100 & partially-inline-libcalls \\
17 & coro-cleanup & 59 & load-store-vectorizer & 101 & post-inline-ee-instrument \\
18 & coro-early & 60 & loop-data-prefetch & 102 & functionattrs \\
19 & coro-elide & 61 & loop-deletion & 103 & mem2reg \\
20 & coro-split & 62 & loop-distribute & 104 & prune-eh \\
21 & correlated-propagation & 63 & loop-fusion & 105 & reassociate \\
22 & cross-dso-cfi & 64 & loop-guard-widening & 106 & redundant-dbg-inst-elim \\
23 & deadargelim & 65 & loop-idiom & 107 & rpo-functionattrs \\
24 & dce & 66 & loop-instsimplify & 108 & rewrite-statepoints-for-gc \\
25 & die & 67 & loop-interchange & 109 & sccp \\
26 & dse & 68 & loop-load-elim & 110 & slp-vectorizer \\
27 & reg2mem & 69 & loop-predication & 111 & sroa \\
28 & div-rem-pairs & 70 & loop-reroll & 112 & scalarizer \\
29 & early-cse-memssa & 71 & loop-rotate & 113 & separate-const-offset-from-gep \\
30 & early-cse & 72 & loop-simplifycfg & 114 & simple-loop-unswitch \\
31 & elim-avail-extern & 73 & loop-simplify & 115 & sink \\
32 & ee-instrument & 74 & loop-sink & 116 & speculative-execution \\
33 & flattencfg & 75 & loop-reduce & 117 & slsr \\
34 & float2int & 76 & loop-unroll-and-jam & 118 & strip-dead-prototypes \\
35 & forceattrs & 77 & loop-unroll & 119 & strip-debug-declare \\
36 & inline & 78 & loop-unswitch & 120 & strip-nondebug \\
37 & insert-gcov-profiling & 79 & loop-vectorize & 121 & strip \\
38 & gvn-hoist & 80 & loop-versioning-licm & 122 & tailcallelim \\
39 & gvn & 81 & loop-versioning & 123 & mergereturn \\
40 & globaldce & 82 & loweratomic & 124 & \texttt{-Oz} \\ 
41 & globalopt & 83 & lower-constant-intrinsics &  &  \\
\end{longtable}
}

\section{The Use of Large Language Models}
We used LLMs to assist in refining the clarity and coherence of the writing in the paper. The LLMs were specifically employed to improve phrasing, ensure academic rigor, and enhance overall readability. Their contribution was strictly in the writing process, and all content was thoroughly reviewed and finalized by the authors. 

\end{document}